\documentclass[useAMS,usenatbib]{mn2e}

\usepackage{graphicx}
\usepackage{txfonts}
\usepackage{natbib}

\title[Modeling SEDs of XTE~J1118+480 and GX~339$-$4]{Constraining Jet/Disk 
	Geometry and Radiative Processes in Stellar Black Holes XTE~J1118+480 
	and GX~339$-$4}

\author[D. Maitra et al.]
{Dipankar~Maitra$^{1}$\thanks{E-mail:D.Maitra@uva.nl}, Sera~Markoff$^{1}$, 
	Catherine~Brocksopp$^{2}$, 
	Michael~Noble$^{3}$, \newauthor Michael~Nowak$^{3}$
	and J\"orn~Wilms$^4$
\\ $^1$ Astronomical Institute ``Anton Pannekoek'', University of Amsterdam, 
        Kruislaan 403, 1098 SJ Amsterdam, The Netherlands
\\ $^2$ Mullard Space Science Laboratory, University  College London,
        Surrey RH5 6NT, UK
\\ $^3$ Kavli Institute for Astrophysics and Space Research,
        Massachusetts Institute of Technology, Cambridge, MA, USA
\\ $^4$ Dr.\ Karl Remeis-Observatory, University of Erlangen-Nuremberg,
        Sternwartstr.~7, 96049 Bamberg, Germany
}

\date{Accepted 2009 April 8. Received in original form 2008 June 9}
\pagerange{\pageref{firstpage}--\pageref{lastpage}}
\pubyear{2009}

\begin{document}
\label{firstpage}

\maketitle

\begin{abstract}

We present results from modeling of quasi-simultaneous broad band (radio
through X-ray) observations of the galactic stellar black hole (BH)
transient X-ray binary (XRB) systems XTE~J1118+480 and GX~339$-$4 using an
irradiated disc $+$ compact jet model. In addition to quantifying the
physical properties of the jet, we have developed a new irradiated disc model
which also constrains the geometry and temperature of the outer accretion disc
by assuming a
disc heated by viscous energy release and X-ray irradiation from the inner
regions. For the source XTE~J1118+480, which has better spectral coverage of
the two in optical and near-IR (OIR) wavelengths, we show that the entire
broad band continuum can be well described by an outflow-dominated model $+$ 
an irradiated disc. The best-fit radius of the outer edge of the disc is 
consistent with
the Roche lobe geometry of the system, and the temperature of the outer edge of
the accretion disc is
similar to those found for other XRBs. Irradiation of the disc by the jet is
found to be negligible for this source. 
For GX~339$-$4, the entire continuum is well described by the jet-dominated 
model only,
with no disc component required. For the two XRBs, which have very
different physical and orbital parameters and were in different accretion
states during the observations, the sizes of the jet base are similar and
both seem to prefer a high fraction of non-thermal electrons in the
acceleration/shock region and a magnetically dominated plasma in the jet. 
These results, along with recent similar results from modeling other galactic 
XRBs and AGNs, may suggest an inherent {\em unity in diversity} in the 
geometric
and radiative properties of compact jets from accreting black holes.

\end{abstract}

\begin{keywords}
         black hole physics---
	 radiation mechanisms: non-thermal and thermal---
	 X-rays: binaries---
	 stars: winds, outflows--
	 accretion, accretion discs--
	 X-rays: individual: (GX~339$-$4, XTE~J1118+480)
\end{keywords}

\section{Introduction}\label{sec:intro}
Soft X-ray transient (SXT) binary systems that occasionally go into  
episodes of high activity or {\em outbursts} are some of the best 
laboratories to study accretion flows near compact ($GM/R\sim c^2$)
objects. Coordinated multi-wavelength campaigns have been carried 
out in recent years 
to observe SXTs simultaneously over a broad frequency range of nine
decades or more from radio to X-rays \citep[see, e.g.,][for a review]
{fender2006}. These campaigns have provided extremely valuable
insights about the accretion in/outflows and helped to constrain the 
theoretical models predicting accretion geometry and physical emission 
processes.
Outbursts of SXTs are usually explained in terms of a disc 
instability model \citep[see, e.g.,][for a review]{l2001}, postulating
a dramatic increase in mass accretion via an accretion disc towards the 
central compact object. The luminosity of a black hole transient (BHT) can 
vary by as much as eight orders of magnitude between quiescence and peak 
outburst, within timescales of weeks to months. Such systems are also ideal for 
studying the formation of jets, which are a by-product of accretion, yet 
poorly understood so far.

Matter spiralling into the gravitational potential well of the compact
object forms an accretion disc, and radiates via dissipation of the
accumulated gravitational potential energy. The observed UV and 
soft X-ray spectra of these sources, during epochs of high mass accretion 
rate, are well modeled by viscous
dissipation from the inner regions of a geometrically thin, optically thick
``Shakura--Sunyaev disc'' \citep{ss1973}.

However in the OIR, the viscously dissipating disc 
model sometimes significantly underpredicts the observed spectral energy 
distribution (SED), and the excess 
is attributed to irradiation of the outer disc by the intense X-ray radiation 
from the inner region close to the compact object. 
This irradiation mechanism has been well explored theoretically  
\citep{c1976,vrtilek1990,kkb1996,d1999,d2001} and their effect observed in 
optical and near-IR wavelengths \citep[e.g.][]{vrtilek1990,vPM1994,m1995,
hynes2002,hynes2005,r2006,mb2008,g2009}.
The most prominent, observable effect of 
irradiation is to heat up the outer regions of the accretion disc to 
temperatures higher than what would be possible solely by viscous dissipation 
of a standard Shakura-Sunyaev disc. This leads to a deviation in the 
shape of the SED of the accretion disc (from that of a standard disc), creating 
a `bump' in the optical through
infra-red (OIR) wavelengths. As the quality of the OIR data improves, the
deviation (i.e. the bump) from  Shakura-Sunyaev spectrum becomes more and 
more apparent, and the need to incorporate irradiation in the model becomes 
important.

In Roche lobe overflowing BHTs where the accretor is 10 times
or more heavier than the donor, X-ray irradiation of the donor is small.
Furthermore irradiation causes the outer disc to flare, shielding most
of the donor star from irradiation \citep{m1978}. While most models assume an
azimuthally symmetric accretion disc, warping of the accretion disc 
\citep[see, e.g.,][and references therein]{od2001} can introduce additional 
uncertainities in shielding the donor from irradiation.

Not all the mass accreted inwards falls onto the compact object. Outflows,
in the form of steady compact jets, or transient, ballistic ejecta are
often seen in compact binary systems \citep[see, e.g.,][for a recent review]
{fender2006}. Steady, compact jets have been observed from an increasing
number of compact binary systems when they are in an {\em X-ray hard state}
(see below). The hallmark signature of such a jet is its
flat to slightly inverted spectrum which, following the classic work by
\citet{bk1979}, is usually attributed to a superposition of self-absorbed
synchrotron emission from segments of a collimated jet. 
For stellar BHTs
the flat spectrum of the jet is expected to extend from radio frequencies
up to IR \citep[$\sim 10^{13-14}$ Hz;][]{markoff2003}, beyond which 
the jet synchrotron becomes optically thin \citep[see, e.g.,][]{cf2002}.
At X-ray frequencies, inverse Compton (IC) upscattering in the jet base 
of synchrotron as well as external photons from the accretion disc
can become comparable and may even dominate the spectrum. As shown in 
\citet{mnw2005}, synchrotron and IC at the jet base can result in spectra 
very similar to those produced by compact Comptonizing coronal 
models \citep{p1998,c1999}. 

During the course of an outburst, the X-ray spectral and temporal features
show an enormous amount of diversity. However extensive observations and 
their analyses over the past few decades have enabled the classification
of spectral and temporal features
into relatively few groups known as {\em X-ray states} \citep[see, e.g.,]
[for extensive definitions and review of X-ray states]{mr2006,hb2005}. 
In particular, two canonical states are often seen:  a ``soft'' state 
characterized by strong thermal emission from the accretion disc and low 
variability in the light curve, and a ``hard'' state characterized by
non-thermal power law emission, high variability and probably a collimated 
outflow. Besides these canonical 
states, a host of intermediate and extreme states with varying ratios of
thermal and non-thermal radiation also are seen. 
Recent OIR observations \citep[e.g.][]{mb2008,russell2008} covering complete 
outbursts of many XRBs have shown that while irradiation usually plays a 
dominant role 
in the evolution of the OIR spectrum during the entire outbursting phase, 
the contribution from a non-thermal source, e.g. a
jet, can become dominant when XRBs (most notably black holes) are in hard 
state. Thus any model which aims to understand disc-jet coupling must 
include irradiation.

We have developed a new spectral model which calculates continuum emission 
from an 
irradiated accretion disc, and in conjunction with a modified version of
the compact jet model developed by \citet{mnw2005}, allows us to constrain
the geometrical and radiative properties not only of the jet, but 
simultaneously those of the accretion disc. The thermal photons from the
irradiated disc are also included in the photon field of the jet for Compton 
scattering, although due to Doppler redshifting in the jet frame, this 
external Compton emission is usually much smaller than the synchrotron self 
Compton (SSC). The new model is fully integrable with the standard spectral 
analysis software ISIS \citep{houck2002} and parallelizable for using in large 
cluster computing environments (also see \S\ref{sec:parallel}).

We have used this new model to analyze broad band quasi-simultaneous 
observations of two galactic BHT systems XTE~J1118+480 and GX~339$-$4.
Simultaneous or quasi-simultaneous 
observations of both sources, during hard X-ray state of their outbursts, have 
revealed broad band continuum emission ranging from radio to X-rays, strongly
suggesting the presence of jets \citep{mff2001,markoff2003,h2005,mnw2005}. 
By studying short time-scale variability in near-IR wavelengths from 
XTE~J1118+480, \citet{h2006} also suggested a non-thermal origin of the
near-IR radiation, with a power law index typical of optically thin synchrotron
emission from a jet.

The known physical properties of the two sources XTE~J1118+480 and GX~339$-$4
have been briefly summarized in \S\ref{sec:data}.
Since analysis of multi-wavelength data requires careful calibration of data 
from various ground-based as well as space-borne instruments, we have also 
presented the 
data reduction and calibration procedures in \S\ref{sec:data}. A brief 
outline of the model, with somewhat more emphasis on the newly introduced 
irradiated disc component, is given in \S\ref{sec:model}, along with a 
description of the fitting procedure. 
The results of our broad band SED modeling suggest that despite the differences
in size, mass, orbital parameters or accretion state, some of the fundamental 
physical parameters characterizing the geometrical and radiative properties 
of the jet are similar for both sources. This similarity may be a global
property of jets formed near compact objects and is discussed in 
\S\ref{sec:discussion}.

\section{Sources, Observations \& Data Reduction}\label{sec:data}
XTE~J1118+480 and GX~339$-$4 are two prototype galactic BHTs with
markedly different properties. 
Both binary systems contain high mass compact objects, 
$8.5\pm 0.6 \rm{M}_\odot$ for XTE~J1118+480 \citep{gelino2006} 
and $>6\ \rm{M}_\odot$ for GX~339$-$4 \citep{munoz2008}, 
thus strongly suggesting for the compact objects to be 
black holes in both cases. 

XTE~J1118+480 lies $62^\circ.3$ north of galactic plane, in the halo. It has
an orbital period of $4.04$ hours \citep{gelino2006}. 
The distance and interstellar column density to XTE~J1118+480 are reasonably 
well known from OIR photometry during quiescence to be $1.72\pm0.1$ kpc and
$1.3\times10^{20}$ cm$^{-2}$ respectively \citep{gelino2006}. 
We adopted a black hole mass of $8.5\ \rm{M}_\odot$ for this source. 
The orbital inclination of XTE~J1118+480, as measured by \citet{gelino2006}
is $68\pm2$ degrees.

GX~339$-$4 lies near the galactic plane, only $4^\circ.3$ south of the galactic 
equator and has an orbital period of $42.14$ hours \citep{h2003}.
We adopted a black hole mass of $7\ \rm{M}_\odot$ for GX~339$-$4 in this work. 
The secondary star in the GX~339$-$4 system has eluded detection so far, thus 
making estimates of masses, distance or line of sight extinction difficult.
Here we have used a distance of 6 kpc and $N_{\rm H}=6\times10^{21}$ 
cm$^{-2}$ for GX~339$-$4, which are consistent with the limits 
given by \citet{h2003,h2004} and \citet{munoz2008}. 

\subsection{OIR and radio data}\label{sec:oirdata}

XTE~J1118+480 went through an outburst during January--February 2005.
A discussion of the OIR light curve morphology of the 2005 OIR outburst of 
XTE~J1118+480 is presented by \citet{z2006}. It was noted by \citet{z2006} 
that the 5--12 keV/ 3--5 keV hardness ratio from the {\em All Sky 
Monitor} \citep[ASM; ][]{levine1996} data onboard the {\em Rossi X-Ray Timing 
Explorer} \citep[RXTE; ][]{brs1993} satellite suggested that the source was
in an X-ray {\em hard} state during the entire outburst of 2005.
For this source we used quasi-simultaneous optical (from the Liverpool 
telescope), near-IR (from UKIRT) and 15 GHz radio data (from the Ryle 
telescope). The full set of multi-wavelength observations obtained during this
outburst of XTE~J1118+480 will be presented in Brocksopp et al. (in prep).

XTE~J1118+480 went through an outburst in 2000 as well. During this outburst
the source was extensively observed by many ground based telescopes as well
as space-borne missions \citep[see e.g.][]{hynes2000,mcc2001,frontera2001,
c2003}. In particular, the source was observed quasi-simultaneously around
2000 April 18th using the Ryle radio telescope, UKIRT, HST, EUVE, Chandra
and RXTE giving an unprecedented broadband spectral coverage. This excellent
data set has been used in several works \citep[e.g.][]{mff2001,esin2001,
yuan2005} to infer properties of the source. We revisit this data set to 
constrain the geometry and radiative properties of both the jet and the
disc using our new disc+jet model.

We used the OIR and radio SED of GX~339$-$4 reported by \citet{h2005}, 
which were obtained during a multi-wavelength campaign to observe its 
outburst of 2002.
For all data sets we used the photometric OIR magnitudes 
and de-reddened them using the interstellar 
extinction law of \citet{ccm1989}. In the absence of definitive measurement
of the ratio of total to selective extinction ($R_V [=A_V/E(B-V)]$) in the
direction of either of the sources, we have taken $R_V=3.1$, the widely 
used value measured by \citet{rl1985} towards the galactic center, and
standardly used for these sources. Thus
it must be kept in mind that if the ``true'' value of $R_V$ is significantly
different from $3.1$ (given neither of the sources are very close to the 
galactic center), it might introduce additional uncertainity in reddening
correction.

\subsection{X-ray data}\label{sec:xraydata}
For both sources, we used data from the
{\em Proportional Counter Array} \citep[PCA; ][]{jahoda2006} and the 
{\em High Energy X-Ray Timing Experiment} \citep[HEXTE; ][]{roth1998} 
onboard the RXTE satellite. 
The X-ray data were downloaded from the NASA HEASARC's public 
archive\footnote{http://heasarc.gsfc.nasa.gov/docs/archive.html} 
and reduced using an in-house script following the standard 
reduction procedures outlined in {\em RXTE cook 
book\footnote{http://heasarc.nasa.gov/docs/xte/recipes/cook\_book.html}} 
using HEASOFT software \citep[v6.1.2][]{arnaud96}. 
To avoid any 
terrestrial contamination, we considered data only for elevation angles
of $10^{\circ}$ or higher. Also any data with pointing offset more than
$0^{\circ}.02$, or within 30 minutes of SAA passage, or trapped electron
contamination ratio more than 0.1 were rejected. Since both the sources
were bright (PCU2 count rate $>40$ counts/s/PCU), the model 
{\em pca\_bkgd\_cmbrightvle\_eMv20051128.mdl} was used for background 
subtraction. Data from all 
the layers of the active PCUs during the observations were
extracted to create the spectra. Systematic uncertainty of 0.5\% was added
in quadrature to the PCA data to account for uncertainties in the 
calibration. For the HEXTE instrument, data from both 
clusters were added. Both PCA and HEXTE data were binned to achieve a 
minimum $S/N$ of 4.5. For the PCA we used the energy range of 3--22 keV 
and for HEXTE 20--200 keV. The normalization of the OIR and radio data was
tied to that of PCA. HEXTE normalization was allowed to vary. 

\section{Model, Analysis and Results}\label{sec:model}
The continuum jet model used in this work is based on \citet{mnw2005},
(henceforth referred to as the {\tt agnjet} model). Given the importance
of irradiation in the observed spectra of XRBs, we have modified the model
to include additional physics (see \S\ref{sec:irrad} for details) that 
allow us to calculate the effect 
of irradiation on the outer disc due to high energy photons from (1) the jet, 
and (2) the inner regions close to the compact object, impinging on the disc. 
We also compute inverse Compton scattering of thermal 
disc photons by the electrons in the jet plasma, although their effect on the
total spectrum is negligible, except near the jet base, due to Doppler 
redshifting of the disc photons in the jet frame.

\subsection{Jet parameters}\label{sec:jetpars} 
The details of the physics of the jet model as well as a full description 
of its main input parameters is given in the appendix of \citet{mnw2005}.
Therefore we only give a brief summary here, and outline the 
modifications made to the \citet{mnw2005} model:

The main parameters that determine the properties of the jet are
the input jet power ($N_j$), 
electron temperature of the relativistic thermal plasma entering at the jet 
  base ($T_e$), 
the ratio of magnetic to particle energy density (a.k.a. the equipartition 
  factor $k$),
physical dimensions of the jet base (assumed to be cylindrical with radius 
  $r_0$ and height $h_0$), and 
the location of the point on the jet ($z_{acc}$) beyond which the a 
  significant fraction of the leptons are accelerated to a power law energy 
  distribution.
$N_j$, parameterized in terms 
of the Eddington luminosity, determines the power initially input into the 
particles 
and magnetic field at the base of the jets.  In the absence of a full 
understanding of 
the mechanism for energizing the jets, $N_j$ plays a similar role to the 
compactness 
parameter in thermal Comptonization models \citep{c1999}.
 A typical value of the number density of leptons ($n$) at the base 
  of the jet (derived from fits) is 
  $n=N_{\rm j}L_{\rm Edd}/(4\gamma_{\rm s}\beta_{\rm s} m_{\rm p} c^3 
		  \pi r_{\rm 0}^2) \sim 10^{14-15}$ cm$^{-3}$. 
Here $L_{\rm Edd}$ is the Eddington luminosity of the source,
$\beta_{\rm s}\sim 0.4$ is the sound speed at the base (see below),
$\gamma_{\rm s}=(1-\beta_{\rm s})^{-1/2}$, 
$m_{\rm p}$ is the proton rest mass, and $r_{\rm 0}\sim 10^7$ cm is
the radius of the jet base, implying an optical depth 
$\tau=n\sigma_T r_{\rm 0}\sim 0.001-0.01$. Since the optical depth is small,
the probability of multiple Compton scattering is very small and we only 
consider single scattering.
For the non-thermal power law electrons, the lower cutoff in the Lorentz
factor is assumed to be equal to peak of the relativistic Maxwell-Juttner 
distribution, i.e. $\gamma_{e;min}=2.23 kT_e/(m_ec^2)$.

Following the prescription of \citet{fb1995,f1996} for maximally efficient
jets, we assume that the bulk speed of the plasma (with adiabatic index 
$\Gamma=4/3$) at the base of the jet is the sound speed given by 
$\beta_{\rm s}=\sqrt{(\Gamma-1)/(\Gamma+1)}\sim0.4$. We do not model the 
particle acceleration in the jet, but just start at the sonic point. Each jet 
then accelerates 
longitudinally along its axis due to pressure gradient and expands laterally 
with its initial sound speed.
The bulk velocity profile, magnetic field strength,
electron density and electron Lorentz factor along the jet are calculated by 
solving the adiabatic, relativistic Euler equation.

In the previous versions of the jet model \citep[e.g. in][]{mnw2005} we used 
the ratio of scattering mean free path to gyroradius of the 
particles in the plasma through which the shock moves ($f_{sc}$), as a model 
parameter.
However given the current lack of physical understanding of the actual shock 
propagation mechanism, in this paper we fixed the shock speed in the plasma 
($\beta_{sh}$) to $0.6$ and used the quantity 
$\epsilon_{sc}=\beta_{sh}^2/$(ratio of scattering mean free path to 
gyroradius) as a model parameter instead of $f_{sc}$. The parameter 
$\epsilon_{sc}$ can be physically interpreted to be a parametrization for the 
shock acceleration rate \citep[$t_{acc}\propto \epsilon_{sc}^{-1}$; see eq.1 
of][]{mff2001}.
The dominant cooling processes within the jet are synchrotron and synchrotron
self-Compton radiation. If the inner edge of the accretion disc is
sufficiently close to the base of the jet, then the soft photons from the
accretion disc, acting as seed photons for external Comptonization, can 
also cool the particles in the jet. 
The jet continuum as observed by an observer on Earth is a superposition 
of spectra from individual segments along the jet axis (taken to be 
perpendicular to the accretion disc plane), calculated by solving the 
radiative transfer equation, after taking into account relativistic 
beaming effects.

\subsection{The irradiated disc}\label{sec:irrad}
Since the coupling between disc and jet is not well understood, we 
assume an independent classical thermal viscous ``Shakura-Sunyaev''  
accretion disc, with a radial temperature profile $T(R)\propto R^{-3/4}$, and
modeled by the parameters $T_{\rm in}$, $r_{\rm in}$, the temperature and 
radius at
the inner edge of the disc. If the outer parts of the disc are irradiated,
then as discussed below, it is further possible to constrain the temperature 
and radius at the outer edge of the disc. Given the mounting 
evidence for
the existence of relativistic outflows from black holes, we also explore the
influence of the jet acting as a source of irradiation heating of the accretion
disc.

\subsubsection{Can jets influence the energetics of the accretion 
disc ?}\label{sec:jet_irrad}
We computed the radiative jet flux incident on the accretion disc as a 
function of radial distance on the disc plane, taking
into account special relativistic beaming effects. In order to obtain an 
{\em upper limit} on the jet-induced heating of the accretion 
disc, we assumed 
that the entire incident radiative flux from the jet $f_{\nu}(R)$ is 
thermalized upon being absorbed by the disc, thus heating the disc locally 
at radius $R$ at a rate $H_{\rm jet}(R)$, given by
\begin{equation}
 H_{\rm jet}(R) = \int_0^\infty f_{\nu}(R) \mbox{ } d\nu
 \label{jet_irrad_eqn}
\end{equation}

We calculated $H_{\rm jet}(R)$ at ten logarithmically spaced radii spanning
from $R_{\rm in}$ to $R_{\rm out}$. The ratio of $H_{\rm jet}(R)$ to the 
local viscous heating ($H_{\rm visc}(R)$) is shown in Fig.~\ref{fig:jet_irrad}.
At large radii ($R>100$ $R_g$), $H_{\rm jet}(R)$
has a $\sim r^{-2.4}$ radial profile, which is somewhat steeper than that 
expected from irradiation by a static corona or the inner accretion disc
(for which the irradiation heating falls as $\sim r^{-1.6{\rm~to~}-2.0}$; 
see, e.g., \S\ref{sec:disc_irrad} and \citet{hynes2005}). However the local
viscous heating rate drops outward even faster 
($H_{\rm visc}(R) \propto R^{-3}$), thus causing the ratio 
$H_{\rm jet}(R)/H_{\rm visc}(R)$ to slowly increase for increasing R.
As the disc radius becomes small and approaches that of
the jet base, the solid angle subtended by the jet increases rapidly causing 
the jet heating to saturate. However the viscous heating continues to 
rise at smaller radii and 
causes $H_{\rm jet}(R)/H_{\rm visc}(R)$ to drop at small radii.
However, as evident from Fig.~\ref{fig:jet_irrad}, even the maximum possible 
jet-induced heating of the disc is about seven orders of magnitude smaller 
than the viscous heating. This confirms that relativistic beaming of the jet
causes a very small fraction of the flux emitted by the jet to fall back on 
the disc, and the resulting influence of the jet on the energetics of the disc
is negligible. 

Light bending 
effects near the black hole can enhance the flux incident on the disc, but by 
not more than a few tens of percent of the numbers calculated above (i.e. by 
assuming Minkowski metric instead of a more realistic Kerr or Schwarzschild 
metric).

\subsubsection{An irradiation source near the compact 
	object}\label{sec:disc_irrad}
Analysis of OIR light curves of XTE~J1118+480 by \citet{z2006} 
suggests there might have been significant contribution of thermal flux
in the optical. The near-IR flux however is reported to be significantly
non-thermal in nature from studies of rapid time variability in the light 
curve, as well as nearly flat IR SED \citep{z2006,h2006}. 
As discussed in the previous paragraph as well as shown in 
Fig.~\ref{fig:jet_irrad},
the jet's contribution in heating the disc is negligible; we
have therefore added an irradiation heating term due to a somewhat more
static source of irradiating X-rays
near the black hole. 
The physical origin of this irradiating source
could be the inner accretion disc near 
the black hole.
Since we assume a radial temperature profile and do not 
solve the local disc structure \citep[which in itself is a detailed 
MHD problem, see, e.g.,][and references therein]{hbk2007}, our model cannot 
discern the source of the irradiating X-rays.

The temperature due to irradiation heating ($T_{\rm irrad}$) is 
usually assumed to 
have a power law radial dependence of the form $R^{-n}$. Depending on 
the initial assumptions about disc structure and geometry of the irradiating
source, theoretical models predict $n$ to be in the range of 0.4-0.5
\citep{vrtilek1990,kks1997,d1999}. 
Given the uncertainities in de-reddening and scarcity 
of data points in the OIR, the exact choice of $n$ is not vital for a 
global understanding of the broad-band SED. We used $n = 3/7$ 
(vertically isothermal disc with disc height $h\propto r^{9/7}$), mainly for 
consistency with earlier work \citep[e.g. by][]
{vrtilek1990,hynes2002,hynes2005}. 

We assume that the total disc heating is a sum of local viscous 
heating, and heating due to a static source of irradiation near the black hole.
The effective temperature of the disc ($T_{\rm eff}$) is therefore related to
the viscous temperature ($T_{\rm visc}$) and the irradiation temperature
($T_{\rm irrad}$) as follows:

\begin{equation}
 T_{\rm eff}^4(R) = T_{\rm visc}^4(R) + T_{\rm irrad}^4(R){\rm ,}
\end{equation}

where $T_{\rm visc} = T_{\rm in}(R/R_{\rm in})^{-3/4}$ and  
      $T_{\rm irrad}= T_{out}(R/R_{out})^{-3/7}$
with $T_{\rm in/out}$, $R_{\rm in/out}$ as free parameters.
Other parameters, viz. masses, inclination and 
distance to the binary system, are taken from values published elsewhere and
kept fixed during the process of fitting.
The radius $R_{\rm irrad}$ where irradiation heating becomes equal to viscous 
heating can be estimated from 
\begin{equation}
 R_{\rm irrad}=\left[ (T_{\rm in}/T_{\rm out})\
	 R_{\rm out}^{-3/7}/R_{\rm in}^{-3/4}\right]^{28/9}.
\end{equation}
Since viscous dissipation dominates for $R < R_{\rm irrad}$, the disc height
at $R_{\rm irrad}$ is estimated from Shakura-Sunyaev's $\alpha$-disc solution 
\citep[see e.g.][]{ss1973, fkr2002} as 
\begin{equation}
 H \ (R_{\rm irrad})=1.7\times 10^8 \ \alpha^{-1/10} \ \dot M_{16}^{3/20} 
	 (M/{M}_\odot)^{-3/8} \ R_{\rm irrad;10}^{9/8} \ f^{3/5}
\label{eq:H_irrad}
\end{equation}

where we have $\dot M_{16}=\dot M/(10^{16} \ {\rm g/s})$,
$R_{\rm irrad;10}=R_{\rm irrad}/(10^{10} \ {\rm cm})$, 
$f=[1-2GM/(c^2 R_{\rm irrad})]^{1/4}$ and chosen $\alpha=0.05$. Similarly,
since $H\propto R^{9/7}$ when $R > R_{\rm irrad}$, the disc height at
$R_{\rm out}$ is given by 
\begin{equation}
H \ (R_{\rm out}) = H(R_{\rm irrad})\times(R_{\rm out} / R_{\rm irrad})^{9/7}.
\label{eq:H_out}
\end{equation}
The solid angle subtended by the outer, irradiation dominated disc as seen from
the compact object is computed from equations~\ref{eq:H_irrad} and 
~\ref{eq:H_out}.

The {\tt agnjet} model computes the
broadband continuum from the jet and the irradiated disc but not 
line emission or reflection features. Therefore
we used an additional {\tt Gaussian} line profile to fit the iron K$\alpha$ 
emission complex near 6.5 keV and the convolution model {\tt 
reflect} \citep{mz1995} for the excess in the range of 10--30 keV, both 
features being 
usually attributed to reflection by the relatively cold accretion disc. 
The column density, taken from elsewhere published values, was fixed during 
the fits. For the {\tt reflect} model, 
which accounts for reflection from neutral material in the accretion disc, 
we used solar abundance and tied the inclination to that of the
disc (and also the jet axis). The energy of the iron line in the 
{\tt Gaussian} model was allowed to vary between 6--7 keV and 
its width was fixed to 0.5 keV. 
The spectral coverage, along with column density ($N_{\rm H}$) towards the 
sources, distances, and masses of the compact objects are given in 
Table~\ref{tab:data}.

\subsection{Model fits}\label{sec:parallel}

Spectral fittings were performed using ISIS \citep[v1.4.9-4; ][]{houck2002}. 
Evaluation of confidence intervals for any physical model is
computationally expensive, and this is true for {\tt agnjet} model as
well.
Therefore we have used the parallelization technique described in
\citet{n2006} and distributed the task of computing
confidence intervals\footnote{Using our own custom versions of the {\tt
cl\_master} and {\tt cl\_slave} scripts described at
http://space.mit.edu/cxc/isis/single\_param\_conf\_limits.html} for $f$
free parameters over $f/2$ or $(f+1)/2$ nodes (depending on whether $f$
is even or odd; each node is an Intel Xeon 3.4GHz processor) of a Parallel 
Virtual Machine \citep{Geist:1994:PPV} running on the LISA cluster in Almere,
Netherlands.
This reduced the runtime to under 24 hours, a greater than 75\%
speedup; it also allowed us to discern finer features in parameter
space by increasing the tolerance resolution by a factor of 500,
while keeping the overall runtime to approximately half the
serial runtime of 96 hours. While these are
welcome improvements, reducing the model runtime even further would enable
us to analyze new data more quickly.  For this reason we are experimenting
with ways of utilizing more than one processor per parameter, such as
partitioning parameter space more finely, and exploring the use of OpenMP
\citep{OpenMP.98} to parallelize internal loops within the model source
code.

Using the distributed parallel computing technique described above, we 
determined the best-fit parameter values as well as their confidence 
intervals. Our confidence intervals correspond to $\Delta \chi^2 = 2.71$ 
for a given parameter (which for {\em normal} distribution would imply 
a $90\%$ single parameter confidence limit).

\subsubsection{XTE~J1118+480 during 2005 outburst}
Of the three data sets presented in this work, the data from the 2005 outburst 
of  XTE~J1118+480 was the faintest and had a steep power law slope in the 
X-rays. The flux in 
the HEXTE bands was extremely low for this data and while regrouping the
data we found that no flux with S/N $>4.5$ could be detected above 60 keV.
Since the single point in
the radio frequency does not constrain the length of the jet, we fixed it to
$10^{13.1}$ cm which is a lower limit on the radiative length of the jet 
assuming that the
radio data point is still on the optically thick synchrotron regime with a 
flat-to-slightly-inverted spectral index. While the fits require a high 
fraction of non-thermal electrons in the post-shock jet ($>0.5$), they are not
very sensitive to the exact value of this fraction, and we fixed its value 
to 0.75. The data cannot be described by a non-thermal jet continuum alone
and an excess in the OIR data requires additional flux which we modeled using
the irradiated accretion disc model described in \S\ref{sec:irrad}. 
This thermal excess in the optical has also been suggested from the OIR 
coverage of the outburst by \citet{z2006} and \citet{h2006}.

The X-ray data (3--60 keV) can be well fit ($\chi^2/\nu=88.8/55$) only with a
power law, however systematic residuals in the 5--7 keV range suggest that a 
weak iron line may be present. Adding a Gaussian line to the power law 
improves 
the fits to $\chi^2/\nu=79.5/52$, with the line center near 6.6 keV and width of
0.1 keV. Given the weak strength of the lines, we fixed the line energy as well
as the width to these values, and did not use any reflection model 
modeling the broad band SED of XTE~J1118+480.

Fits performed by allowing $r_{\rm in}$ and $T_{\rm in}$ of the accretion disc 
to vary freely results in good fits with quite small values of both
$r_{\rm in}$ and $T_{\rm in}$. 
There are uncertainities in our understanding of the physical properties of the 
innermost regions of the disc close to ISCO, dereddening of interstellar extinction, and also uncertainities in instrumental response and calibration. These 
concerns make detection of a cold disc with small $r_{\rm in}$, 
$T_{\rm in}$ and its physical origin open to debate 
\citep{mhm2006,miller2006,rykoff2007,rs2007,g2008,dangelo2008,reis2009}.
Another issue with the XTE~J1118+480 data is the jet/disc
inclination. While we usually fix the inclination to a value published
elsewhere, the extremely flat radio to IR spectrum of XTE~J1118+480 requires a
much smaller inclination than the published value of $\sim 70^\circ$ 
\citep[see, e.g.,][]{gelino2006,c2003,m2001,z2002}. Allowing the inclination to 
vary for this source results in jet inclination of $\sim25^\circ$.

Therefore we tried two different models for this data set of XTE~J1118+480. 
In the first model (Model 1; second column in Table~\ref{tab:fits}), apart from 
other regular parameters described above, the jet/disc inclination as well as 
$r_{\rm in}$ and $T_{\rm in}$ were allowed to vary freely. In the second, 
somewhat more conservative model (Model 2; third column in 
Table~\ref{tab:fits}), the jet inclination was fixed to $30^\circ$, 
$r_{\rm in}$, $T_{\rm in}$ were also fixed to $30$ $R_g$ and $0.1$ keV
respectively. The best-fit values for the second model also gives 
similar jet parameters as the first model, although with a slightly higher
jet power and steeper electron distribution index. Also the second model 
favors a somewhat smaller and hotter outer disc edge.

\subsubsection{XTE~J1118+480 during 2000 outburst}\label{sec:j1118_2000}
We used the broadband SED data of XTE~J1118+480 obtained around 2000 April 18, 
published in \citet{mcc2001} and \citet{hynes2000}. 
While analyzing this data set a broad dip in the
combined EUVE and Chandra spectra was noted between 0.15--2.5 keV (see e.g. 
the residuals in Fig.~\ref{fig:j1118_2000fit}) which cannot be modeled using 
the standardly 
used column density for this source ($1.3\times 10^{22}\ 
{\rm cm}^{-2}$). This feature was been noted by previous works as well, and is 
attributed to metal absorption in a partially ionized gas, 
\citep[see e.g.][]{esin2001}.
We have therefore excluded the energy range between 0.15--2.5 keV 
from our broadband fitting. Results of fitting indicate that during this
observation emission from the jet dominates the radio and IR regions. 
Contribution from the accretion disc starts becoming dominant in the optical.
However, as discussed in \citet{hynes2000}, the EUVE fluxes are 
extremely sensitive to the extinction law, and the slope of the dereddened 
EUVE SED can range between $+2 \geq \alpha \geq -4$ 
($F_{\nu}\sim\nu^{\alpha}$) for the permissible range of N$_{\rm H}$. 
Here we have used the EUVE fluxes reported by \citep{hynes2000, mcc2001} 
and a column depth of N$_{\rm H}=1.3\times10^{20}$ cm$^{-2}$, which results 
in a very steep EUVE slope. This cutoff in the EUVE data would imply a cold,
truncated inner disc with with r$_{\rm in}\sim 340$ R$_g$ and 
T$_{\rm in}\sim 2.9\times 10^5\ {\rm K}$.
Given the low temperature of the inner disc and
the consequent absence of a irradiating soft X-ray photons, it is 
not unexpected that we do not find any signature of irradiation in this
case (i.e. T$_{\rm out}$ is not constrained). As in the case of the 2005 data
of this source, the X-ray emission in this model is entirely dominated by
optically thin synchrotron emission from the jet. While the fits require a 
small nozzle ($h_0/r_0 \sim 0.2-0.6$), they are not very sensitive to the 
exact value of this parameter, and we fixed its value to 0.4. We note that
this value is abot a factor of three smaller than what is found typically 
in other sources, or even during the 2005 outburst of this source.

\subsubsection{GX~339$-$4 during 2002 outburst}\label{sec:gx339}
The X-ray data of GX~339$-$4 cannot be well fit by a power law alone. 
Using simple power law+ Gaussian model gives unacceptable fit to the data
with reduced $\chi^2$ of 5.9. Convolving a power-law+gaussian with a 
reflection model give better fits, but still with rather a large 
reduced $\chi^2$
of 2.3. We found that the GX~339$-$4 data can be fitted well without including
any thermal component, and the entire broadband SED is well described by
the superposition of synchrotron and inverse Compton photons from the 
pre-shock region near the jet base as well as optically thick synchrotrons
from post-shock regions farther from the jet base. 
From the radio-IR slope, the required inclination is between 45--50 degrees, 
but not well constrained and hence was fixed at 47 degree. 
The length of the jet was fixed to $10^{14.2}$ cm and fraction of non-thermal
electrons fixed to 0.75 for the same reasons given for XTE~J1118+480. Since the 
spectrum of GX~339$-$4 could be entirely dominated by jet emission, and the 
source was
in a bright, hard X-ray state \citep{h2005}, the input jet power is much 
larger than that of XTE~J1118+480. In fact the jet power for this observation 
is larger than any of the previous observations of this source 
\citep{markoff2003,mnw2005}. Also the acceleration region starts much farther
out for the GX~339$-$4 data set. 

If the temperature of the thermal electrons ($T_{\rm e}$) at the base is left 
completely free, as in ``Model 1'' (fifth column in Table~\ref{tab:fits}), 
then $T_{\rm e}$ for the best-fit model is comes out
to be rather high, and suggests that the SED at energies $>10$ keV is almost
entirely due to SSC emission from the base (see top two panels in 
Fig.~\ref{fig:gx339fit}). 
The geometry of the base is rather compact and the source is in a bright hard 
state; therefore a strong SSC emission could imply that the photon density
at the base becomes high enough for pair processes to become important. Since 
pair production is not included in the model, an estimate of the importance of 
pair processes for the best fit models were done as follows: 
the pair production 
rate ($\dot n_{pp}$) was calculated from the photon field at the base of the 
jet using the angle averaged pair production cross section 
\citep{gs1967,bs1997}; pair annihilation rate ($\dot n_{pa}$) was calculated 
following \cite{svn1982}. For ``Model 1'' of GX~339$-$4, 
$\dot n_{pp}\sim 10^{17}$ cm$^{-3} s^{-1}$ and 
$\dot n_{pa}\sim 10^{12}$ cm$^{-3} s^{-1}$, suggesting that pair processes
could be important in this case. Therefore we
explored the parameter space, limiting the electron temperature to be less 
than $5\times 10^{10}$ K and somewhat less compact base (larger r$_{\rm 0}$,
h$_{\rm 0}$) to search for solutions which also satisfy the condition 
$\dot n_{pp}<\dot n_{pa}$ (see ``Model 2'', sixth column in 
Table~\ref{tab:fits} and bottom panels in Fig.~\ref{fig:gx339fit}). 
The fit derived number density ($n$), $\dot n_{pa}$ and $\dot n_{pp}$ for the 
different models presented in Table~\ref{tab:fits}, are given in 
Table~\ref{tab:pairs}.

In addition to the continuum jet model, an additional Gaussian line centered 
at 6.3 keV and reflection from the accretion disc is needed to 
obtain good fits for the GX~339$-4$ data. Presence of the line and reflection
suggests the presence of an accretion disc. However, as in the case for 
XTE~J1118+480, the disc is cold and therefore not detected by the PCA.

The best-fit model parameters, associated confidence intervals, $\chi^2$ 
statistic and corresponding chance probabilities for both 
sources are shown in Table~\ref{tab:fits}.
Given the simplicity of the model it is hardly unexpected that the 
reduced $\chi^2$ values obtained from the fitting are not very close to 1 in 
some cases. 
For instance, the viscous+irradiated disc model
does not include details of disc atmosphere or ionization balance equations, 
and is therefore not able to predict the spectral breaks in the
optical data. Because of its simplicity, this is not meant to be a real
model of the disc, although it incorporates the dominant ongoing physical 
processes and can reproduce the general shape of the broadband continuum.
Moreover, while fitting a data set spanning ~10 decades in frequency (or 
energy) space, the long lever arm between radio and X-rays constrains the
jet model more strongly than the overall chi-squared \citep[see e.g.][]{
markoff2008}.
Best fit models along with data and residuals are shown in 
Figures~\ref{fig:j1118_2005fit}, 
~\ref{fig:j1118_2000fit} and~\ref{fig:gx339fit}.

\section {Discussion and Conclusions}\label{sec:discussion}
We have analyzed broad band multi-wavelength observations of the galactic
black hole transient systems XTE~J1118+480 and GX~339$-$4, during their 
outbursts in 2000 and 2005 for XTE~J1118+480, and during the 2002 outburst of
GX~339$-$4. The conclusions from our study are summarized below.

 The data for XTE~J1118+480 cannot be fit well with a jet continuum model 
 alone, due to an excess in the OIR fluxes. During these observations, in OIR 
 the source was at least 3.5 
 magnitude brighter than quiescence; therefore the contribution from 
 the M1 V secondary star \citep[][ and references therein]{c2003} is small.
 Modeling the data from the 2000 outburst and 2005 outburst of this source
 we found that the outer disc was strongly irradiated during the 2005 data.
 It appears that the inner edge of the accretion disc was much closer to the
 central black hole during the observations of 2005, thus providing an ample
 source of soft X-ray photons to irradiate the outer disc. 
  On the other hand, the sharp cutoff in the EUVE data and absence of any 
  signature of irradiation from the outer disc suggests a recessed, cold 
  accretion disc for the 2000 data. However, as discussed in 
  \S\ref{sec:j1118_2000}, the slope of the EUVE data 
  depends very strongly on the assumed column depth, and a small decrease in 
  N$_H$ can flatten the EUVE SED significantly \citep[see e.g.][]{hynes2000}.
  Such a flat EUVE SED would be expected from a disc with small r$_{\rm in}$ 
  and comparatively higher T$_{\rm in}$ \citep{reis2009}.

\begin{figure*}
\includegraphics[width=0.75\textwidth, angle=-90]{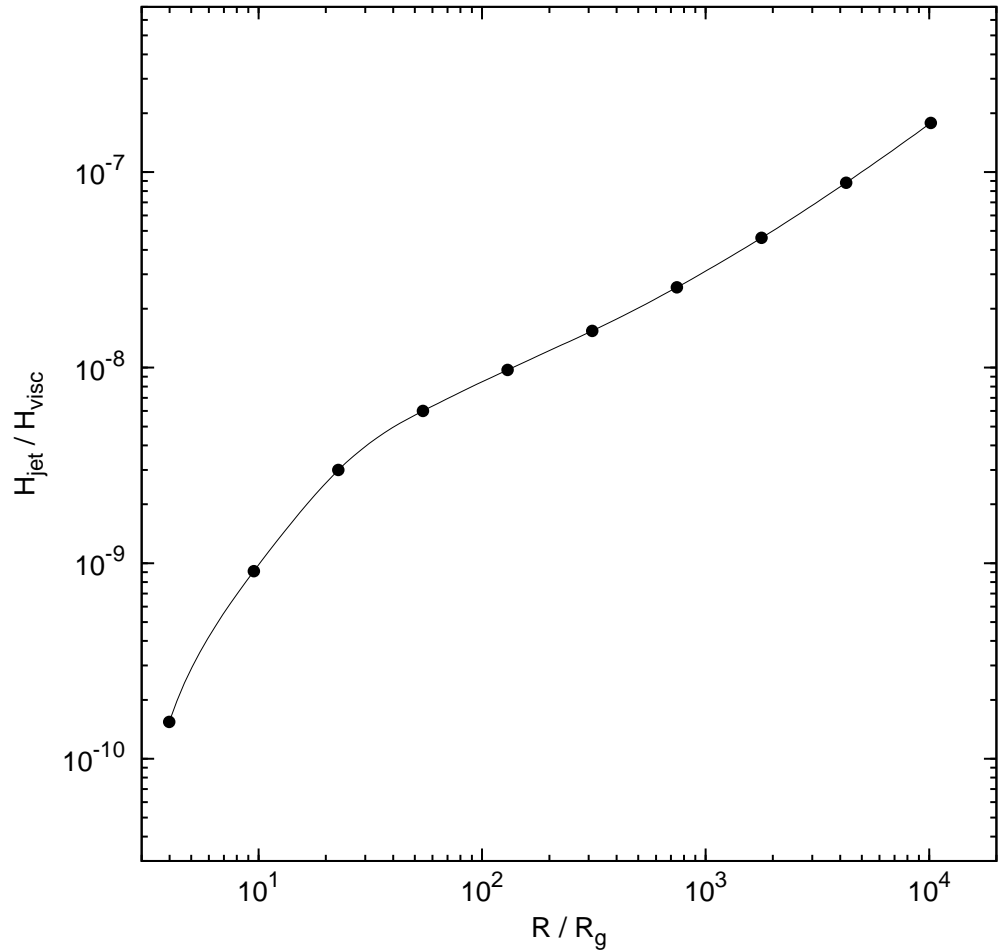}
\caption{Comparing viscous heating ($H_{\rm visc}$) and {\em maximum 
possible} ``jet-heating'' ($H_{\rm jet}$) for XTE~J1118+480. 
The filled
circles are the computed upper limits of the ratio of jet induced 
heating to viscous heating ($H_{\rm jet}/H_{\rm visc}$), as discussed in 
\S\ref{sec:jet_irrad}, joined by
a smooth spline. $T_{\rm in}$, $R_{\rm in}$ and $R_{out}$ for the disc were 
determined from broadband ``Model 1'' fits to the SED. 
For $R>100\ R_g$, the jet induced irradiation heating term falls more
slowly ($\sim R^{-2.4}$) than the viscous heating ($\sim R^{-3}$), which
causes the ratio $H_{\rm jet}/H_{\rm visc}$ to slowly increase at increasing
radii. For $R_{\rm in} < r_0$, the jet heating saturates but the viscous 
heating 
continues to increase for smaller radii, causing $H_{\rm jet}/H_{\rm visc}$
to drop sharply for decreasing radii.
\label{fig:jet_irrad}}
\end{figure*}

\begin{figure*}
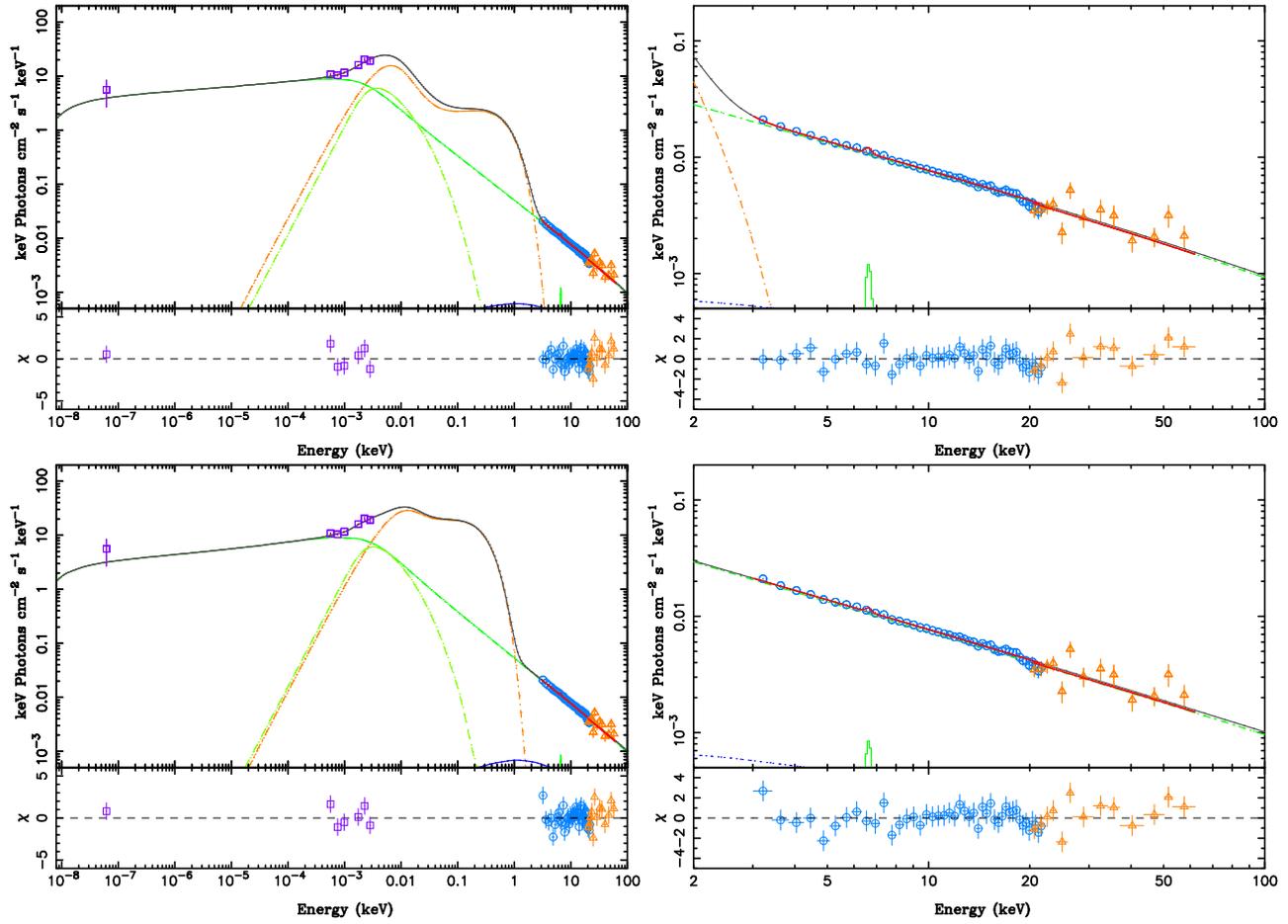

\includegraphics[width=0.48\textwidth]{plot_all_j1118_1.ps} 
\includegraphics[width=0.48\textwidth]{plot_xray_j1118_1.ps} 
\includegraphics[width=0.48\textwidth]{plot_all_j1118_2.ps} 
\includegraphics[width=0.48\textwidth]{plot_xray_j1118_2.ps} 
\caption{Jet model fits and residuals for the XTE~J1118+480 data
from its 2005 outburst.
Left panels show the entire broad band SED from radio through X-rays, and a 
zoom of the X-ray region is shown in the right panels.
Radio and IR data from the Ryle telescope and UKIRT respectively are
shown by purple squares, RXTE/PCA data are shown by by blue circles and 
RXTE/HEXTE data by orange triangles.
The top panels are for ``Model 1'' in Table~\ref{tab:fits} (second column) 
where along with other parameters (see text), the jet inclination, 
$r_{\rm in}$ and $T_{\rm in}$, were also allowed to vary.
The value of the reduced chi-squared for best fit parameters of ``Model 1'' 
is 1.41.
The bottom panels are for ``Model 2'' in Table~\ref{tab:fits} (third column) 
where jet inclination=$30^\circ$, $r_{\rm in}$=$30$ $R_g$ and 
$T_{\rm in}$=$0.1$ keV.
The value of the reduced chi-squared for best fit parameters of ``Model 2'' 
is 1.69.
The dark-green dash-dotted curve shows the post-shock synchrotron 
contribution, the green dashed curve shows the pre-shock synchrotron, blue
dotted curve shows the Compton upscattered component. Flux from the irradiated 
accretion disc is shown by the orange dash-dotted curve.
The solid grey line
shows the total jet+disc model continuum spectrum without convolving through 
detector responses, interstellar extinction, the iron line or reflection. 
The iron line near 6.6 keV is shown by the thick, solid green line. 
The red line shows the properly forward-folded models taking
into account detector responses, interstellar extinction, the iron line and 
reflection. Since interstellar extinction is small for XTE~J1118+480, and the 
iron line and reflection features are very weak too, the forward-folded model 
(red line)
is almost indistinguishable from the jet+disc model continuum (solid grey) in
this figure as well as in Fig.~\ref{fig:j1118_2000fit}. However for GX~339$-$4
(Fig.~\ref{fig:gx339fit}), where the iron line as well as reflection features 
are stronger, and the extinction is also larger, the difference between the 
unfolded and folded models (grey and red lines) becomes apparent.
\label{fig:j1118_2005fit}}
\end{figure*}

\begin{figure*}
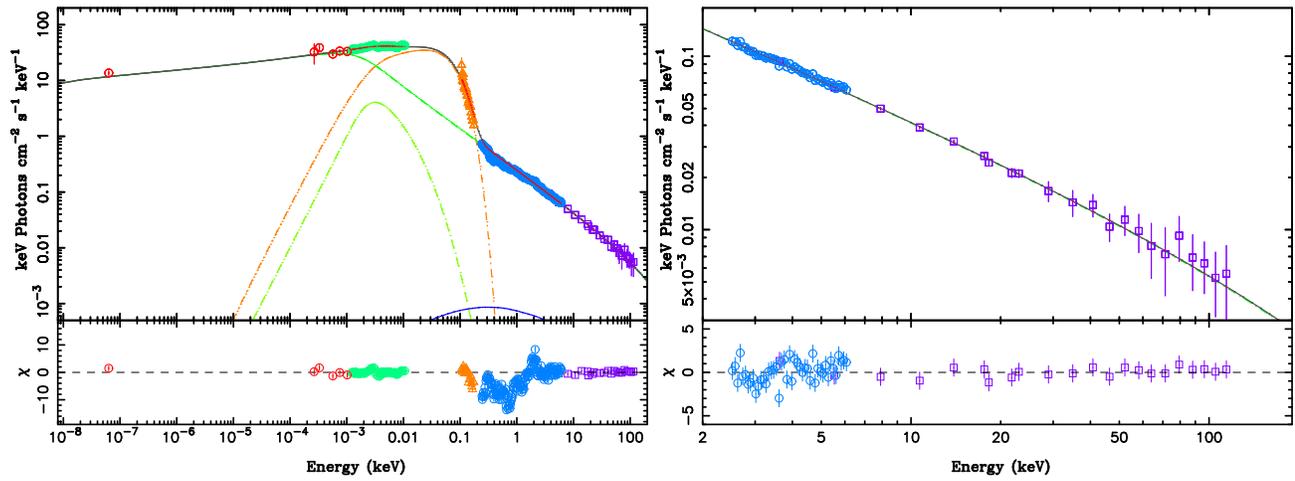

\includegraphics[width=0.48\textwidth]{plot_all_2000.ps} 
\includegraphics[width=0.48\textwidth]{plot_xray_2000.ps}
\caption{Jet model fits and residuals for the XTE~J1118+480 data
from its 2000 outburst \citep[taken from][]{mcc2001} are shown in the 
left panel.
The radio data from the Ryle telescope and IR data from UKIRT are shown by red
circles, HST spectrum using green, EUVE data using orange triangles, Chandra 
spectrum using blue circles and RXTE data using purple squares.
Note the dip in the 0.15--2.5 keV region which is attributed to the presence
of a warm absorber \citep{esin2001}. This region was excluded from our fits.
A zoom of the X-ray region after excluding 0.15--2.5 keV data is 
shown in the right panel.
The colour coding for the model components are the same as in 
Fig.~\ref{fig:j1118_2005fit}.
\label{fig:j1118_2000fit}}
\end{figure*}

\begin{figure*}
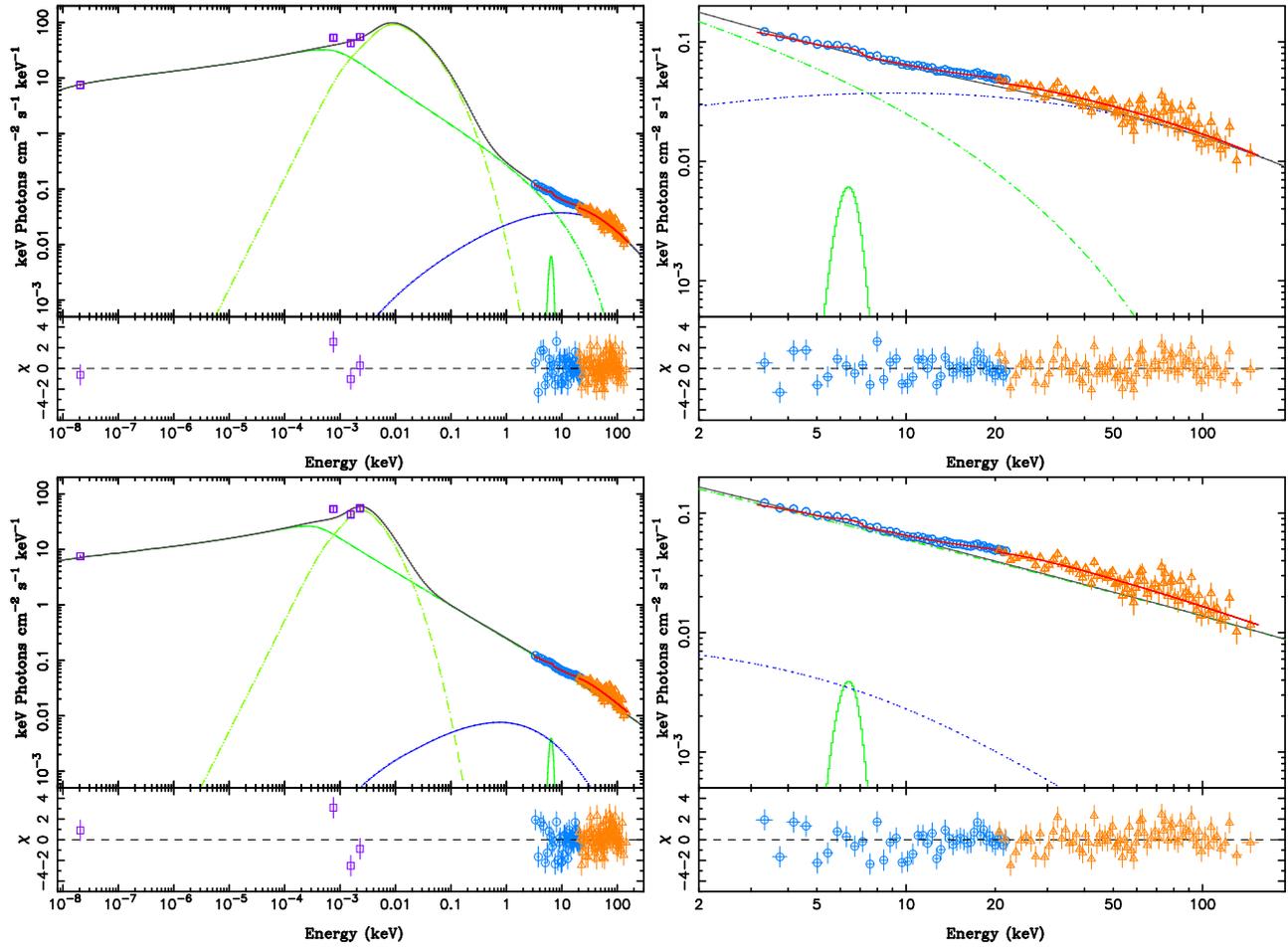

\includegraphics[width=0.48\textwidth]{plot_all_gx339_1.ps}
\includegraphics[width=0.48\textwidth]{plot_xray_gx339_1.ps}
\includegraphics[width=0.48\textwidth]{plot_all_gx339_2.ps}
\includegraphics[width=0.48\textwidth]{plot_xray_gx339_2.ps}
\caption{Same as Fig.~\ref{fig:j1118_2005fit}, but for the source
GX~339$-$4. 
The models presented here are purely jet models without any 
contribution from an accretion disc.
The top panels are for ``Model 1'' (Table~\ref{tab:fits}; fifth column) where
along with other parameters the temperature of the thermal electrons at the
base of the jet ($T_{\rm e}$) was allowed to vary freely. The reduced 
chi-squared for this model is 1.22.
Note the relatively strong inverse Comptonization component in the X-rays, 
originating at the jet base due to SSC emission in this model, compared to 
that in the case of XTE~J1118+480. 
The bottom panels are for ``Model 2'' (Table~\ref{tab:fits}; sixth column) 
where we constrained $T_{\rm e}$ to be less than $5\times10^{10}$ K, reducing
the SSC emission at the base. The base is less compact and acceleration starts
farther out along the jet for ``Model 2'' compared to ``Model 1''.
The reduced chi-squared for this second model is 1.40.
While the first model is a statistically better fit than the second, the photon
density at the base can become high enough for pair processes to become
important in the first model (see \S\ref{sec:gx339},\S\ref{sec:discussion} and 
Table~\ref{tab:pairs}), which is not the case for the second model.
Also note the strong iron line near $6.3$ keV and the 
reflection ``bump'' near $\sim20$ keV, suggesting
the presence of a cold accretion disc not detected at energies to which 
RXTE is sensitive.
\label{fig:gx339fit}}
\end{figure*}

 Our calculations in \S\ref{sec:jet_irrad} show that at any point on the
 surface of the accretion disc, the heating of the disc by the jet is seven or
 more orders smaller than the viscous heating. 
 There are two factors which largely reduce the amount of jet radiation that 
 is incident on the disc: Firstly Doppler de-beaming - since the bulk motion
 of the jet is relativistic, special relativistic Doppler beaming concentrates
 most of the emergent radiation from the jet in the direction of its motion,
 i.e. away from the disc. A second, but significant factor that also 
 contributes to the dimunition of jet flux incident on the disc is the fact 
 that the post-shock jet synchrotron spectra peaks at increasingly longer 
 wavelengths at increasingly larger distances from the base of the jet,
 thus creating fewer X-ray photons that can heat the disc.
 This is illustrated in 
 Fig.~\ref{fig:jet_irrad} where we show the ratio of jet-heating to viscous 
 heating as a function of radius. 
 Since heating by the jet alone cannot 
 produce the OIR excess in Fig.~\ref{fig:j1118_2005fit} we surmised that 
 this the OIR excess is indeed caused by the more conventional form of 
 irradiation, i.e. irradiation by soft X-ray photons from the 
 inner-disc.
 From an energetics point of view, it is easy to see that the 
 inner accretion disc alone (see Fig.~\ref{fig:j1118_2005fit}, at around 1 keV) 
 radiates at least two orders or more energy than the outer region. 
 Furthermore, it is well known that the outer edge of the accretion disc 
 can flare-up or warp \citep[e.g.][]{od2001}, thus making the outer 
 parts of the disc more easily visible from the inner regions.
 
 Fits to the 2005 data of XTE~J1118+480 using the jet model+irradiated disc 
(described in \S\ref{sec:disc_irrad})
 suggest a cold disc ($T_{\rm in}\sim 0.2\ keV$) with inner radius close 
 to 
 the innermost stable circular orbit ($6$ $R_g$ for a Schwarzschild black hole,
 and smaller if spinning). Given the uncertainities associated with using a
 simple $T\sim R^{-3/4}$ model which ignores correction factors like 
 spectral hardening \citep{st1995}, full disc structure
 as well as proper boundary conditions at the inner edge \citep{z2005}, the 
 errors associated with the estimates of the inner disc parameters are most 
 likely an underestimate. 
 Another source of uncertainity comes from that
 fact that we assume the jet to be perpendicular to the disc. A misaligned
 jet \citep{mac2002}, as seems likely for XTE~J1118+480, could introduce 
 additional error in the determination of $r_{\rm in}$.
 Along with the uncertainities in instrument calibration, modeling and 
 jet-disc alignment,
 the sharp low energy cutoff of PCA sensitivity below $3$ keV makes detection 
 of a cold disc with small $r_{\rm in}$ even more difficult than a recessed cold
 disc. Nevertheless the numbers for $r_{\rm in}$ and $T_{\rm in}$ we obtained 
 for XTE~J1118+480 are similar to those obtained by 
 several others for black hole transients in their hard state
 \citep[e.g.][]{disalvo2001,mhm2006,miller2006,rs2007,rykoff2007,reis2009}. 
 If such a cold disc with small r$_{\rm in}$ is indeed realised, then 
 Compton scaterring of the disc photons could be strongly anisotropic, with 
 most of the scattered emission beamed toward the disc \citep{hp1997}. This 
 could compensate partly the beaming effects due to the relativistic 
 motion of the gas, and increase the illumination of the disc (both 
 irradiation and reflection effects). Also in this case there could be a
 radiative feedback due to the local illumination of the disc by the base of 
 the jet which is similar to the situation in standard accretion disc 
 corona models \citep{hm1993} which would act as a thermostat keeping 
 the temperature at the base lower than what is assumed here.
 A somewhat
 more conservative model, where the temperature and radius of the inner edge
 of the accretion disc were fixed to $0.1$ keV and $30$ R$_{\rm g}$ respectively
 (Model 2 in Table~\ref{tab:fits}) also describes the data well, showing the
 difficulty in detecting a cold disc with small r$_{\rm in}$ from the available
 data.

 The best-fit outer radii of XTE~J1118+480 for both {\em Model 1} and {\em 
 Model 2} are smaller than the circularization radius, using the latest 
 values of orbital parameters published by \citet{gelino2006}. These results
 are therefore consistent with the assumption that 
 accretion is occuring via Roche lobe overflow. The observation was made
 during the decline from outburst peak. Hence the discrepancy between the
 circularization radius and $R_{out}$ inferred from fits, in the context of
 standard disc instability models, could be due to an inward moving cooling 
 wave. A decreasing trend in the outer {\em radiative disc radius} was also
 observed by \citet{hynes2002} for the system XTE~J1859+226.
 However the value of $R_{out}$ in our case depends largely on the five OIR 
 data points,
 all of which lie on the Rayleigh-Jeans region of the irradiated disc. Given
 the plethora of spectral features seen in OIR spectra of X-ray binaries in
 outburst \citep[see, e.g.,][]{hynes2005,cc2006}, the limitation of 
 using a few bands is obvious.

 Analysis of the data from the 2000 outburst of XTE~J1118+480 by 
 \citet{esin2001} and 
 \citet{yuan2005} showed that the UV and higher energy data can be explained 
 by an ADAF model. The flux from the ADAF model alone however falls sharply at 
 energies below optical and underestimates IR and any lower energy data (e.g. 
 radio), thus requiring an additional jet component \citep{yuan2005,nm2008}. In 
 the jet-ADAF model \citep{yuan2005}, the X-ray photons have a thermal 
 Comptonization origin, whereas in the 
 current work, the X-ray photons can originate from either synchrotron (e.g in
 XTE~J118+480), or inverse Compton scattering, or a combination of both 
 (e.g. in GX~339$-$4). Better simultaneous broadband coverage and future
 X-ray polarization studies can help to estimate the relative importance 
 of these two emission components which dominate the X-ray spectra of
 X-ray binaries in hard state.

 From the available data, the X-ray spectrum of XTE~J1118+480 during both 
 outbursts show little or no curvature at energies higher than $\sim10$ keV. 
 Reflection features like the iron line complex and the reflection ``bump'' 
 near 20 keV are also extremely weak. 
 Detailed analysis of the X-ray spectra taken by Chandra and RXTE, during 
 the 2000 outburst of this source by \citet{miller2002} also showed extremely 
 weak reflection features, as would be expected for a jet+recessed disc 
 scenario.
 From a phenomenological point of view, since the
 curvature is small, using a broken power law to describe data of X-ray
 binaries in hard state has been shown to work remarkably well, e.g., for
 GX~339$-$4 by \citet{nowak2005} and for Cyg~X$-$1 by \citet{wilms2006}.
 Residuals of a power law fit to the X-ray continuum in the 2005 data set of
 XTE~J1118+480
 show small systematic variations near 5--7 keV, which 
 improves slightly upon addition of a Gaussian line. However, given the 
 weakness of the line, its energy or width cannot be constrained well. 
 Total contribution from various source of inverse 
 Comptonization (SSC within the jet base and external 
 Comptonization of soft thermal photons from the accretion disc) is also 
 much smaller compared to that of GX~339$-$4. 
 The X-ray region of the SED of XTE~J1118+480 is dominated by post-shock
 synchrotron photons for both 2000 and 2005 outbursts.
 We note that the electron acceleration region seems to be in the nozzle 
 of the jet in the case of 2005 outburst data of XTE~J1118+480 while it is 
 farther out in the jet in GX~339$-$4 as well as the 2000 outburst of 
 XTE~J1118+480. While exhaustive searching of the $\chi^2$ space confirms that
 this is required by the data, it is not clear why the acceleration region is
 inside the nozzle for the 2000 data of XTE~J1118+480. Another unexpected 
 result is that the base of the jet seems to be quite small ($h_0/r_0\sim0.4$)
 for the 2000 outburst data of XTE~J1118+480, whereas $h_0/r_0\sim1.5$ for
 other outbursts of XTE~J1118+480 as well as other sources (see e.g. 
 \citet{mnw2005,gallo2007,mig2007}). Given the uncertainties in our 
 understanding of jet formation and the acceleration region, and the
 simplicity of the model, here we report the best fit parameters as obtained.
 The particle energy distribution index ($p$) of the 
 relativistic post-shock electrons for data from both outbursts of 
 XTE~J1118+480 is $\sim$ 2.5--2.6. This value is
 much steeper than what could be expected from an initial distribution of 
 particles accelerated by a diffusive shock process  \citep[$\sim 1.5-2$;]
 []{hd1988}. However it is possible that for XTE~J1118+480 the observed 
 X-ray bandpass is at a higher energy than the cooling break energy, where 
 the particle distribution steepens from $p$ to $p+1$ due to enhanced 
 cooling. One of the key assumptions in our model is time independence of
 particle distributions. However if the steep
 power law drop-off of the observed photon index is indeed due to enhanced
 cooling, or more generally a departure from the assumed equillibrium
 between the heating and cooling processes, a more thorough time dependent
 analysis of particle evolution is neccessary. We will present time dependent
 particle evolution in the context of jets in a forthcoming paper.

 The jet inclination of XTE~J1118+480 is still a matter of concern.
 The optically measured system inclination is near $70^{\circ}$
 \citep{gelino2006,c2003,m2001,z2002}. However, given the flat radio-to-IR 
 fluxes, jet inclinations higher than $\sim30^{\circ}$ would be very difficult 
 to fit which might suggest a large misalignment between between the disc 
 and the jet \citep{mac2002}. Evidence of misalignment between disc and jet 
 has been seen in several stellar X-ray binary systems, e.g., GRO~J1655$-$40 
 \citep{gbo2001,hr1995}, SAX~J1819$-$2525 \citep{h2000,o2001}, XTE~J1550$-$564 
 \citep{h2001,o2002}, as well as in several AGNs, e.g., NGC 3079 
 \citep{k2005}, NGC 1068 \citep{cap2006} and NGC 4258 \citep{cap2007}.
 Allowing the inclination to vary for XTE~J1118+480 during the 
 fits gives a best-fit inclination of $\sim25^{\circ}$, and it is almost 
 impossible to find statistically good fits (e.g. with $\chi^2/\nu < 2$) 
 for inclinations larger than $30^{\circ}$. Previous fits by \citet{mff2001} 
 also required a similar, small inclination. In our model we assume the lateral
 expansion and longitudinal acceleration of the jet to be driven solely via the
 solutions of the relativistic Euler equation for adiabatic expansion. If this
 is not the case, and e.g. the jet radius and acceleration are determined by 
 some other process(es) leading to a stronger magnetic field and/or smaller 
 bulk velocity along the jet, that could lead to a flatter radio-IR spectrum
 as well \citep[also see the discussion in \S3.3.3 of][]{mig2007}.

 The data for GX~339$-$4 does not formally require a thermal component (e.g. an
 accretion disc or emission from the secondary star) and the entire broad
 band emission from radio to X-rays is satisfactorily well fit by 
 non-thermal photons originating from pre- and post-shock synchrotron, and
 synchrotron self Compton emission mainly from the jet base. 
 Previous fits to the 3--100 keV X-ray data by \citet{h2005} also suggested
 that the disc contribution within this energy range is $\sim 1\%$ of the total
 luminosity or less.
 However 
 presence of the line near 6.5 keV and signature of reflection suggests
 a cold accretion disc near the black hole which cuts off exponentially at an 
 energy significantly below the lower detection limit of the RXTE/PCA.
 Since the X-ray spectrum has a luminosity of $\sim 5\%$ L$_{\rm Edd}$, the
 disc luminosity is presumably $\sim 0.05\%$ L$_{\rm Edd}$ or smaller.
 
 The data for GX~339$-$4 were taken when the source was in a bright, 
 X-ray hard state \citep{h2005}. The jet power $N_j$ is much higher than 
 that obtained by \citet{mnw2005}, who studied data from previous outbursts 
 of the same source. We also obtain a higher equipartition factor ($k$) and 
 shallower particle distribution index. It is interesting to note that 
 $r_0$ and $z_{acc}$ for GX~339$-$4 are always found to be higher than 
 those found for other sources. 
 If the electron temperature at the base is
 left unconstrained (e.g. ``Model 1'' in the fifth column of 
 Table~\ref{tab:fits}), then the density of photon field from 
 best-fit model is high enough that pair processes may become important. 
 In this case many photons are up-scattered above the pair production 
 threshold, and the rate at which pairs are produced is much larger than the
 rate at which they are annihilated, thereby changing the density at the base 
 of the jet. Constraining the electron temperature to be less than 
 $5\times 10^{10}$ K, we 
 present another fit (``Model 2'', sixth column of Table~\ref{tab:fits}) where 
 the photon density at the base is small and
 pair production rate is much smaller than the pair annihilation rate. For this
 fit the the fractional pair annihilation rate $(\dot n_{\rm pa}/n)$ is also 
 much smaller than unity, so that pair processes are not important for this fit.

 Using the irradiated disc+jet model presented in this paper we have shown 
 that in certain cases (e.g. the 2000 outburst data of XTE~J1118+480) there 
 are not enough X-ray photons from the inner regions to irradiate the disc 
 appreciably. The 2005 outburst data of XTE~J1118+480 
 presents an intermediate case where the contribution from an irradiated outer 
 disc is comparable to the jet emission in OIR. The 2002 data of GX339$-$4 
 presents the extreme case where the entire broadband SED is dominated by 
 emission from the jet.

 The jet model has been used to analyze broad band SEDs of the stellar mass 
 black holes XTE~J1118+480 \citep[][this work]{mff2001}, GX~339$-$4 \citep[]
 [this work]{mnw2005}, Cyg~X-1 \citep{mnw2005}, A0620-00 \citep{gallo2007},
 GRO~J1655-40 \citep{mig2007}, as well as that of the supermassive black holes
 in the nucleus our own galaxy \citep[Sgr A*;][]{m2007} and that of 
 M81 \citep{markoff2008}. From the analysis of the spectra of these sources
 we note that the geometry of the jet base, scaled in units of gravitational 
 radius, does
 not vary largely. The radius of the jet base, $R_0$, lies in the
 range of 10--100 $R_g$, and the ratio of $h_0/r_0$ at the nozzle lies
 within 0.4--11, both suggesting a relatively compact jet base and the 
 universal nature of mass scaling in accretion onto compact objects. 
 The fraction of particles accelerated to a power law distribution beyond
 $z_{acc}$, while not very well constrained by the fits, preferes a value
 in the range of 0.6--0.9 and is fixed to 0.75. Considering that most of the
 known acceleration mechanisms in astrophysics do not have such a high 
 efficiency, this number seems rather high and requires further exploration.
 The particle acceleration index ($p$) lies within 2.2--2.7, which is steeper 
 than that expected from a relativistic shock \citep[1.5--2, see, e.g.]
 {hd1988}, suggesting that in these cases the X-ray energies may lie beyond 
 the {\em cooling break} \citep{k1962}, so that the power law index of 
 observed SED steepens by an additional amount of 0.5.
 
 The magnetic-to-particle energy densities at the jet base ($k$) is often 
 found to be greater than unity. MHD simulations of relativistic jet 
 launching \citep{mg2004,nm2004,devilliers2005,fm2008} suggest that near the 
 jet base, close to the launching points, the jets are mostly Poynting flux 
 dominated and have small plasma fractions (ratio of gas pressure to magnetic 
 pressure is less than unity). We are focusing on inner regions where this is 
 likely still to be the case.  In GRBs and AGN, most data is only relevant to 
 regions well beyond the magnetosonic fast points, where the flow has been 
 driven towards equipartition via conversion of magnetic energy to particle 
 energy.

 Temperature of the thermal electrons at the jet base ($T_e$) is in the range 
 of $2-7\times10^{10}$ K for the stellar black holes and $\sim 10^{11}$ K for 
 the supermassive black holes. The flow near the jet base could be a 
 radiatively inefficient accretion flow (RIAF), i.e. an ADAF 
 \citep{narayan1995,rees1982,shapiro1976}, or in light of the the magnetic 
 domination, an MDAF \citep{meier2005}. In such a case, the ion temperature 
 can be as high as the virial temperature, and electrons at $~10^{9-10}$ K 
 \citep{shapiro1976,nym1996}. Keep in mind that for an outflow model, the 
 particles are also not required to be gravitationally bound, and processes 
 in the jet launching can heat the particles beyond virial, in theory.

Most likely the similarity in the derived 
physical parameters among the diverse sources and their diverse accretion 
states is not coincindental, and may hint towards an inherent similarity in 
the process of formation and propagation of compact jets.
The results presented above show the importance of quasi-simultaneous,
multi-wavelength, broad band SEDs in constraining geometrical and radiative 
parameters associated with accertion flows near compact objects.

\vspace{5mm}
\emph{Acknowledgements}.
This work was supported mainly by Netherlands Organisation for Scientific 
Research (NWO) grant number 614000530. JW acknowledges DLR grant 50OR0701. 
We would like to thank Guy Pooley for helping with the radio data, Michelle 
Buxton and Mickael Coriat for the OIR data reduction.
We would also like to thank the anonymous referee for his/her constructive
criticisms that have greatly improved the paper. 
DM thanks Asaf Pe'er for a discussion on pair processes in plasmas.
It is a pleasure to acknowledge the hospitality of International Space 
Science Institute (ISSI) in Bern where a significant amount of this work 
was carried out.

This research has made use of data 
obtained from the High Energy Astrophysics Science Archive Research 
Center (HEASARC), provided by NASA's Goddard Space Flight Center. This 
work has also made use of the United Kingdom Infrared Telescope 
(UKIRT) which is operated by the Joint Astronomy Centre on behalf of 
the Science and Technology Facilities Council of the U.K.
The Liverpool Telescope is operated on the island of La Palma by 
Liverpool John Moores University in the Spanish Observatorio del Roque 
de los Muchachos of the Instituto de Astrofisica de Canarias with 
financial support from the UK Science and Technology Facilities Council.
We have also used radio data from the Ryle telescope operated by
Mullard Radio Astronomy Observatory.
Numerical computations for this work were done on the LISA workstation
cluster, which is hosted by Stichting Academisch Rekencentrum Amsterdam (SARA) 
computing and networking services, Amsterdam.

\begin{table*}
\begin{tabular}{llll}
\hline
\hline
 & XTE~J1118+480 (2000 outburst)
 & XTE~J1118+480 (2005 outburst)
 & GX 339$-$4    (2002 outburst) 
 \\
\hline
MJD of observation
 & 51652
 & 53393
 & 52367-68
 \\
Spectral Coverage & & & \\
   Radio: 
   & 15 GHz 
   & 15 GHz 
   & 5 GHz 
 \\
   UV,Opt/IR: 
   & M, L, K, H, J-bands, EUVE and HST spectra 
   & K, H, J, V, B-band 
   & H, I, V-band 
 \\
   X-ray: 
   & Chandra (0.24--7 keV), RXTE (3--110 keV) 
   & RXTE (3--70 keV) 
   & RXTE (3--200 keV) 
 \\
$N_{\rm H}$ ($10^{22}\ cm^{-2}$) 
 & 0.013 
 & 0.013 
 & 0.6 
 \\
$M_{\rm BH}$ (${M}_\odot$) 
 & 8.5 
 & 8.5 
 & 7 
 \\
Distance (kpc) 
 & 1.7 
 & 1.7 
 & 6 
 \\
\hline
\end{tabular}
\caption{Spectral coverage and source parameters used for the fits.
For the 2000 outburst of XTE~J1118+480 the radio data were obtained
using the Ryle telescope, IR using the UKIRT, optical spectra using 
the HST, UV spectra using EUVE, X-ray spectra using Chandra and 
RXTE. The SED was constructed from data published in \citet{mcc2001}.
For the 2005 outburst of XTE~J1118+480 the radio data were obtained 
using the Ryle telescope,
IR using the UKIRT and optical using the Liverpool telescope. 
The column density, mass of the black hole and distances were taken 
from the recent observations by \citet{gelino2006} for the source 
XTE~J1118+480. For GX~339$-$4, the SED was constructed from data published 
in \citet{h2005}, black hole mass is lower limit by \citet{munoz2008}, 
distance and column density from the ranges given by \citep{h2004}.}
\label{tab:data}
\end{table*}

\begin{table*}
 \begin{tabular}{llllll}
 \hline
 \hline
  Parameter (Unit) & 
  XTE~J1118+480    & 
  XTE~J1118+480    & 
  XTE~J1118+480    & 
  GX~339$-$4       &
  GX~339$-$4       \\
   & 
  (2005 outburst; Model 1) & 
  (2005 outburst; Model 2) &
  (2000 outburst)          & 
  (2002 outburst; Model 1) &
  (2002 outburst; Model 2)
   \\
 \hline
 $N_{\rm j}$ ($10^{-3}$ ${\rm L}_{\rm Edd}$) & 
 $ 2.32 _{-0.06}^{+0.02}$      & 
 $ 2.77 _{-0.05}^{+0.01}$      & 
 $ 7.2 _{-0.5}^{+0.1}$         & 
 $ 55 _{-9}^{+16}$             & 
 $ 83.8 _{-0.2}^{+0.2}$          
 \\
 & & & & \\
 $r_0$ (R$_{\rm g}$)           & 
 $ 9.39 _{-0.17}^{+0.18}$      & 
 $ 11.8 _{-0.2}^{+0.1}$        & 
 $ 20.8 _{-1.4}^{+1.0}$        & 
 $ 33 _{-2}^{+3}$              & 
 $ 114 _{-11}^{+7}$              
 \\
 & & & & \\
 $h_0/r_0$                     & 
 $ 1.589 _{-0.002}^{+0.013}$   & 
 $ 1.39 _{-0.02}^{+0.03}$      & 
 $0.4\star$                    & 
 $ 1.7 _{-0.1}^{+0.2}$         & 
 $ 1.51 _{-0.03}^{+0.02}$        
 \\
 & & & & \\
 $T_{\rm e}$ ($10^{10}$ K)     &
 $ 4.22 _{-0.02}^{+0.11}$      & 
 $ 4.04 _{-0.02}^{+0.01}$      & 
 $ 3.76 _{-0.07}^{+0.09}$      & 
 $ 6.7_{-0.5}^{+0.4}$          & 
 $ 3.8 _{-0.2}^{+0.01}$          
 \\
 & & & & \\
 $z_{\rm acc}$ (R$_{\rm g}$)   &
 $ 12.6 _{-0.0}^{+1.5}$        & 
 $ 12.6 _{-0.0}^{+2.8}$        & 
 $ 8.2 _{-0.2}^{+3.3}$         & 
 $ 252 _{-130}^{+212}$         & 
 $ 526 _{-41}^{+85}$             
 \\
 & & & & \\
 $p$                           & 
 $ 2.63 _{-0.01}^{+0.01}$      & 
 $ 2.71 _{-0.01}^{+0.01}$      & 
 $ 2.48 _{-0.01}^{+0.02}$      & 
 $ 2.28 _{-0.03}^{+0.03}$      & 
 $ 2.20 _{-0.01}^{+0.02}$        
 \\
 & & & & \\
 $\epsilon_{\rm sc}$ ($10^{-3}$)  &   
 $ 9.9 _{-0.1}^{+0.02}$        & 
 $ 29.9 _{-3.0}^{+0.3}$        & 
 $ 5.5 _{-0.8}^{+1.4}$         & 
 $ 0.16 _{-0.08}^{+0.08}$      & 
 $ 44 _{-1}^{+2}$                
 \\
 & & & & \\
 $k$                           & 
 $ 1.9 _{-0.1}^{+0.1}$         & 
 $ 2.3 _{-0.2}^{+0.1}$         & 
 $ 2.1 _{-0.1}^{+0.1}$         & 
 $ 2.2 _{-0.8}^{+1.4}$         & 
 $ 2.38 _{-0.01}^{+0.01}$        
 \\
 & & & & \\
 Inclination (deg)             &
 $ 25.1 _{-0.0}^{+0.1}$        & 
 $30\star$                     & 
 $30\star$                     &
 $47\star$                     & 
 $47\star$                       
 \\
 & & & & \\
 $r_{\rm in}$ (R$_{\rm g}$)    & 
 $ 2.6 _{-0}^{+5}$             & 
 $ 30\star$                    & 
 $ 341 _{-22}^{+4}$            & 
 Not constrained               & 
 Not constrained                 
 \\
 & & & & \\
 $T_{\rm in}$ (keV)            &
 $ 0.254_{-0.008}^{+0.004}$    & 
 $ 0.1\star$                   & 
 $0.025_{-0.002}^{+0.001}$     & 
 Not constrained               & 
 Not constrained                 
 \\
 & & & & \\
 $r_{\rm out}$ (R$_{\rm g}$)   &
 $ 15700 _{-350}^{+160}$       & 
 $ 8850 _{-1550}^{+1800}$      & 
 $ 29400 _{-5750}^{+610}$      & 
 Not constrained               & 
 Not constrained                 
 \\
 & & & & \\
 $T_{\rm out}$ (K)             &
 $ 18100 _{-600}^{+1000}$      & 
 $ 30500 _{-100}^{+200}$       & 
 $ 650 _{-550}^{+2150}$        & 
 Not constrained               & 
 Not constrained                 
 \\
 & & & & \\
 $f \ (10^{-4})$               &
 $1.2$                         & 
 $1.0$                         & 
 Not constrained               & 
 Not used                      & 
 Not used                        
 \\
 & & & & \\
 $E_{\rm Line}$ (keV)          & 
 $6.63\star$                   & 
 $6.63\star$                   & 
 Not used                      &
 $ 6.3 _{-0.2}^{+0.2}$         & 
 $ 6.3 _{-0.2}^{+0.3} $          
 \\
 & & & & \\
 $W_{\rm Line}$ (eV)           & 
 $39.3$                        & 
 $27.7$                        & 
 Not used                      &
 $93.5$                        & 
 $58.7$                          
 \\
 & & & & \\
 $\Omega/2\pi$                 &
 Not used                      & 
 Not used                      & 
 Not used                      &
 $ 0.22 _{-0.14}^{+0.10}$      & 
 $ 0.35 _{-0.02}^{+0.02}$        
 \\
 & & & & \\
 $\chi^2/\nu$ [Q]            & 
 $70/50$      [0.032]        & 
 $90/53$      [0.0011]       & 
 $140/137$    [0.406]        & 
 $137/112$    [0.054]        & 
 $157/112$    [0.0032]         
 \\
 \hline
 \end{tabular}
\caption{Best fit parameters for XTE~J1118+480 and GX~339$-$4, and 
$\Delta \chi^2 < 2.71$ confidence intervals. 
A star ($\star$) next to a number indicates that the parameter was 
frozen to that value.
$N_{\rm j}$ = jet normalization; 
$r_{\rm 0}$ = nozzle radius;
$h_{\rm 0}/r_{\rm 0}$ = height-to-radius ratio at base;
$T_{\rm e}$ = pre-shock electron temperature;
$z_{\rm acc}$ = location of acceleration region along the jet;
$p$ = spectral index of post-shock electrons; 
$\epsilon_{\rm sc}$ = 0.36/ratio of scattering mean free path to gyroradius 
	          (see \S\ref{sec:jetpars});
$k$ = ratio between magnetic and electron energy densities;
Inclination = orbital inclination as seen from Earth (also the jet 
		inclination);
$r_{\rm in/out}$ = radius of the inner/outer edge of the irradiated accretion 
	disc;
$T_{\rm in/out}$ = temperature at the inner/outer edge of the irradiated 
	           accretion disc. While $T_{out}$ is in Kelvin, $T_{\rm in}$ 
		   is reported in $keV$ as is customary for X-ray observers;
$f$ = $(2\pi)^{-1}\times$ the solid angle subtended by the outer, irradiation
      dominated disc as seen from the inner disc; 
$E_{\rm Line}$ = energy of the iron K$\alpha$ line complex from the {\tt 
	gaussian} model;
$W_{\rm Line}$ = equivalent width of the line;
$\Omega/2\pi$ = reflection fraction from the {\tt reflect} model.
$\chi^2/\nu$ is the reduced chi-squared and $Q$ is the corresponding chance
probability.
For fitting the data of XTE~J1118+480 from 2000, the energy range
of $0.15-2.5$ keV was excluded \citep[see text, and also][]{esin2001}. Also the
disc inner edge parameters (r$_{\rm in}$, T$_{\rm in}$) are extremely sensitive
to the extinction column depth (\S\ref{sec:j1118_2000}).
}
\label{tab:fits}
\end{table*}

\begin{table*}
 \begin{tabular}{lccc}
 \hline
 Source/Outburst/Model & 
 log ($n$ [cm$^{-3}$] ) & 
 log ($\dot n_{pa}$ [cm$^{-3}$ s$^{-1}$] ) & 
 log ($\dot n_{pp}$ [cm$^{-3}$ s$^{-1}$] ) \\
 \hline
 XTE~J1118+480/2005/1 & 13.9 & 12.0 & 7.6 \\
 XTE~J1118+480/2005/2 & 13.8 & 11.8 & 6.8 \\
 XTE~J1118+480/2000 & 13.7 & 11.7 & 5.8 \\
 GX~339$-$4/2002/1 & 14.2 & 12.4 & 17.1 \\
 GX~339$-$4/2002/2 & 13.4 & 11.0 & 6.1 \\
 \hline
 \end{tabular}
 \caption{The lepton number density ($n$), pair annihilation rate
 ($\dot n_{pa}$), and pair production rate ($\dot n_{pp}$) at the base of the
 jet for the various models presented in Table~\ref{tab:fits}.}
 \label{tab:pairs}
\end{table*}


\clearpage

\clearpage

\label{lastpage}

\end{document}